**Possible Upper limits on Lorentz Factors in High Energy Astrophysical Processes**

C. Sivaram and Kenath Arun

Indian Institute of Astrophysics, Bangalore

**Abstract:** Gamma-ray bursts (GRBs) are the most luminous physical phenomena in the universe. The relativistic effect on the blast wave associated with the GRB introduces the gamma factor. Here we put an upper limit on the gamma factor via constraints on maximal power allowed by general relativity and hence set upper limits on other observable quantities such as deceleration distance. Also upper limits are set on the high energy particle radiation due to constraints set by cosmic microwave background radiation.

GRBs consist of flashes of gamma rays that last from seconds to hours, and it is now believed that the long-duration bursts (> 2 sec) are associated with the beamed energy from a specific kind of hypernova event. [1, 2]

During the explosion a blast wave is generated which pushes the fragments at very high speeds. The energy associated with the blast waves is given by.

$$E = mc^2 = \left(\frac{4}{3}\pi R^3\right)(nm_P)c^2 \qquad \text{... (1)}$$

In the case of gamma ray bursts, considering the relativistic effects, we have: [3, 4]

$$E = \frac{4}{3}\pi(ct)^3 nm_P c^2 \Gamma^8 \qquad \text{... (2)}$$

Where, $\Gamma = \frac{1}{\sqrt{1 - v^2/c^2}} \approx 100$

The size of the region, in the relativistic case, is given by: $R = \Gamma^2 ct$, which contributes a factor of $\Gamma^6$. The ambient number density in the observer's frame is also enhanced by a factor of $\Gamma$, and the propagation of the blast wave within the bulk motion contributes another factor of $\Gamma$, effectively increasing the energy by a total of $\Gamma^8$.

For a burst time scale of ~10s, ambient number density n ~$10^3$/cc, and $\Gamma \approx 100$ (typically), equation (2) gives:

$$E \approx 10^{51} ergs \qquad \text{... (3)}$$

And the corresponding power is given by:

$$\dot{E} \approx 10^{50} ergs/s \qquad \text{... (4)}$$

When a spherical object of mass *M* collapses under gravity to a black hole, the time scale is given by: $GM/c^3$. Then the corresponding maximum power (Gunn luminosity) (allowed by general relativity) is given by: [5]

$$\dot{E}_{\max} = \frac{Mc^2}{GM/c^3} = \frac{c^5}{G} \approx 10^{59} ergs/s \qquad \text{... (5)}$$

From equations (2) - (5) an upper limit on the gamma factor of the blast wave turns out to be: $\Gamma < 1000$

This may give the highest possible $\Gamma$ factor, i.e. the bulk Lorentz factor for the relativistic process for such blasts. Similarly, other quantities observed will also be constrained. For instance, deceleration distance by a factor of $\Gamma^{-2/3}$ and time by $\Gamma^{-8/3}$, i.e. by a factor of $\frac{1}{100}$ and $\frac{1}{10^8}$ respectively.

As these bursts are also believed to produce high energy particles and gamma rays, their energies would also be similarly constrained. For example, the highest energy photons are supposed to be produced by synchrotron self Compton process, i.e. SSC. If the particles', in this case electrons, Lorentz factor is $\gamma$, then the energy of the photons is boosted by $\gamma^4$, i.e., we have for the photon energy: [6, 7]

$$E_{SSC} \propto (\gamma^4 \omega_B)\Gamma \qquad \text{... (6)}$$

Where, $\Gamma$ is the bulk Lorentz factor, $\omega_B \left( = \frac{eB}{2\pi m_e c} \right)$ is the gyrofrequency (in the rest frame)

The energy photons are cut-off by the CMB, through: $\gamma_{HE} + \gamma_{CMB} \rightarrow e^+ + e^-$

This gives an upper limit of $10^{16}$ eV at present epoch. The bulk motion of the CMB also boosts up $\gamma_{CMB}$ by another factor of $\Gamma$, so the cut-off for high energy is given by:

$$\sqrt{E_{HE}\Gamma E_{CMB}} \sim m_e c^2 \qquad \text{... (7)}$$

That is a cut-off of ~10TeV.

For $\Gamma \sim 10^3$ and for a magnetic field of $B \sim 10^{-4} G$, this would produced by electron of energy:

$$E_{elec} = \gamma m_e c^2 \sim 3 \times 10^{11} eV \qquad \text{... (8)}$$

where $\gamma \sim 5 \times 10^5$

At higher redshifts the energy will correspondingly scale with z as:

$$\sqrt{E_{HE} \Gamma E_{CMB} (1+z)} \sim m_e c^2 \qquad \text{... (9)}$$

For z ~10, and $\Gamma \sim 10^3$, $(E_{HE})_{cut-off} \sim 1\text{TeV}$

Therefore the maximum energy of gamma rays seen at present epoch from such objects would be ~100GeV.

The highest energy cosmic ray particles, which may be gamma photons, are observed to be ~$10^{21}$ eV. This gives the cut-off energy of ~$10^{18}$eV. This, from equation (6), would imply $\gamma \sim 7 \times 10^6$

Again, from GR, the highest individual particle energy is ~Planck energy, again given by quantum gravity effects. This is $\sim \left( \frac{\hbar c^5}{G} \right)^{1/2} \sim 10^{28} eV = 10^{19} GeV$, [8] which again with the additional $\Gamma$ factor (due to bulk motion of CMB) gives the cut-off energy as ~$10^{16}$GeV.

So if this were to be a gamma ray, the maximum particle Lorentz factor $(\gamma_{\max})$, together with the upper limit $(\Gamma_{\max})$ would imply: $\gamma_{\max} \sim 4 \times 10^8$.

The energy lost by the electron at a redshift of z is given by:

$$\dot{E} \approx c \sigma_T \gamma^2 a T^4 (1+z)^4 \qquad \text{... (10)}$$

The temperature T is also boosted by $\Gamma$. Therefore we have:

$$\dot{E} \approx c \sigma_T \gamma^2 a (\Gamma T)^4 (1+z)^4 \qquad \text{... (11)}$$

For electrons with energy given by equation (8), the energy loss becomes:

$\dot{E} \approx 6$ ergs/s, for $T = 3K$ and $z = 10$, and maximal $\Gamma = 10^3$ (which follows from equations (1), (2) and (5))

And the corresponding life time is given by:

$$\tau = \frac{E}{\dot{E}} \propto \frac{1}{\gamma \Gamma^4 (1+z)^4} \qquad \text{... (12)}$$

Which for the above $\dot{E}$ gives:

$$\tau \approx 0.1s \qquad \text{… (13)}$$

For the highest individual particle energy (following from equation (6)), the energy loss becomes:

$\dot{E} \approx 3 \times 10^6$ ergs/s

Corresponding to a life time of:

$$\tau \approx 10^{-4} s \qquad \text{… (14)}$$

These limits thus puts constraints on the source sizes emitting such particles.